# A Study of $B^+ \to p\bar{p}K^+$ and a Search for a $\Theta^{*++}$ Pentaquark Candidate in $B$ Decay.

Tetiana Berger-Hryn'ova

*Stanford Linear Accelerator Center,
MS 61, 2575 Sand Hill Road, Menlo Park, California 94025, USA*

For the BABAR Collaboration

A study of the decay $B^+ \to p\bar{p}K^+$ is performed using 81 $fb^{-1}$ of data collected at the $\Upsilon(4S)$ with the BABAR detector at PEP-II. The branching fraction of $B^+ \to p\bar{p}K^+$ is measured to be $(6.7 \pm 0.9 \pm 0.6) \times 10^{-6}$. An upper limit on the branching fraction of $B^+ \to \Theta^{*++}\bar{p}$, where $\Theta^{*++}$ is a narrow state decaying to $pK^+$, is set to be $1.5 \times 10^{-7}$ for $1.43 < m(\Theta^{*++}) < 1.85 \, \text{GeV}/c^2$ at 90% confidence level. All results are preliminary.

## 1. Introduction

This paper describes a measurement of the branching fraction(b.f.) of the charmless baryonic three-body decay $B^+ \to p\bar{p}K^+$ and a study of its resonant structure[1]. The current value[2] of the b.f. for this mode is $(5.7^{+0.7}_{-0.6} \pm 0.7) \times 10^{-6}$. This mode is interesting as it can be used to search for an $I = 1$, $I_3 = 1$ pentaquark $\Theta^{*++}$ in the decay $B^+ \to \Theta^{*++}\bar{p} \to p\bar{p}K^+$, where $\Theta^{*++}$ would be a member of the baryon 27-plet with quark content $uuud\bar{s}$[3]. The $\Theta^{*++}$ mass has been predicted[4] to lie in the region $1.43 - 1.70 \, \text{GeV}/c^2$. This analysis is performed using 89 million $B\bar{B}$ pairs collected at the $\Upsilon(4S)$ resonance with the BABAR detector[5] at PEP-II.

## 2. Analysis Method

The kaon and proton identification(PID) is based on $dE/dx$ information from the tracking systems for $p_T < 0.7 \, \text{GeV}/c$ or the measured Cherenkov angle and the photon yield observed in the Cherenkov detector(DIRC) for $p_T > 0.7 \, \text{GeV}/c$. The $B^+ \to p\bar{p}K^+$ signal events are isolated using the kinematic constraints of $B$ mesons produced at the $\Upsilon(4S)$: $m_{ES} = [(E_{cm}^2/2 + \mathbf{p}_i \cdot \mathbf{p}_B)^2/E_i^2 - \mathbf{p}_B^2]^{1/2}$ and $\Delta E = E_B^* - E_{cm}/2$, where $E_{cm}$ is the total center-of-mass(cm) energy, $(E_i, \mathbf{p}_i)$ is the four-momentum of the initial $e^+e^-$ system and $\mathbf{p}_B(E_B^*)$ is the $B$-candidate momentum(energy) in the laboratory(cm) frame. In the cm frame, continuum events are jet-like while $B\bar{B}$ events are spherical. Four event-shape variables[6] are combined into a Fisher discriminant in order to suppress continuum background. The main sources of $B\bar{B}$ background are the $b \to c\bar{c}s$ transitions, where $B^+ \to X_{c\bar{c}}K^+$,

Work supported in part by the Department of Energy Contract DE-AC02-76SF00515

Presented at DPF 2004:
Annual Meeting of the Division of Particles and Fields (DPF) of the American Physical Society (APS),
August 26-31, 2004, Riverside, CA

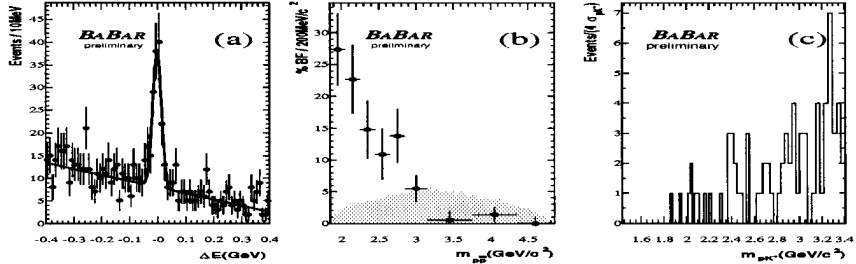

Fig. 1. (a) $\Delta E$ distribution of on-peak data (5.27< $m_{ES}$ < 5.29 GeV/$c^2$ and $m_{p\bar{p}}$ <2.85 GeV/$c^2$, 3.128<$m_{p\bar{p}}$<3.315 GeV/$c^2$ and $m_{p\bar{p}}$>3.735 GeV/$c^2$); (b) The b.f. as a function of $m_{p\bar{p}}$: points are on-peak data, the shaded region is phase-space signal Monte Carlo; (c) The $m_{pK^+}$ distribution (not efficiency-corrected) for data events in the pentaquark signal region.

$X_{c\bar{c}} = \eta_c$, $J/\psi$, $\psi(2S)$, $\chi_{c0,1,2}$ and $X_{c\bar{c}} \to p\bar{p}$ ("charmonium" background). The expected $B\bar{B}$ non-charmonium background is less than one event.

## 3. Results

### 3.1. Branching fraction measurement

A fit to the $\Delta E$ projection (5.27<$m_{ES}$<5.29 GeV/$c^2$) of data outside the charmonium region (see Fig. 1(a)) with a first order polynomial and a single Gaussian yields 114±14 signal events. To calculate the b.f. the $\Delta E$ projections (5.27<$m_{ES}$<5.29 GeV/$c^2$) for the bins of $m_{p\bar{p}}$ weighted by the efficiency for reconstructing the signal in each of the bins (excluding 2.85<$m_{p\bar{p}}$<3.15GeV/$c^2$ as discussed below), are summed; the total distribution is fitted as described above and scaled by the number of charged $B$ mesons in the sample($N_B$): $\mathcal{B}(B^+ \to p\bar{p}K^+) = (\sum \frac{N_{m_{p\bar{p}}}}{\epsilon_{m_{p\bar{p}}}})/N_B$, where $N_{m_{p\bar{p}}}$ is the number of events and $\epsilon_{m_{p\bar{p}}}$ is the efficiency in each of the $m_{p\bar{p}}$ bins. The resulting b.f. is $\mathcal{B}(B^+ \to p\bar{p}K^+; m_{p\bar{p}}<2.85\text{GeV}/c^2) = (5.4 \pm 0.7) \times 10^{-6}$. To measure the total b.f. an estimate of the expected number of charmonium events, otherwise indistinguishable from signal, is needed. The number of the combinatoric background events in 2.85<$m_{p\bar{p}}$<3.15GeV/$c^2$ region is estimated from the $\Delta E$ fit to be 18.0±2.6 events. The fit of the first order polynomial (for the signal and combinatoric background contributions) and two single Gaussians for the $\eta_c$ and $J/\psi$ mass peaks to the $m_{p\bar{p}}$ distribution in the 2.85<$m_{p\bar{p}}$<3.15GeV/$c^2$ region yields 28.5 ± 3.1 signal and combinatoric events. Thus the signal contribution in this region is 10.5 ± 4.0 events. To subtract the contribution from the other charmonium modes the PDG[7] values for the appropriate b.f. are used to estimate their yields. The total b.f. is $(6.7 \pm 0.9) \times 10^{-6}$. The observed $p\bar{p}$ mass spectrum (Fig. 1(b)) differs from a phase-space distribution; it peaks dramatically at low $p\bar{p}$ mass.

### 3.2. $\Theta^{*++}$ pentaquark search[6]

For a $\Theta^{*++}$ search only events with 5.276 < $m_{ES}$ < 5.286 GeV/$c^2$, |$\Delta E$|<0.029 GeV and $m_{pK^+}$<3.4 GeV/$c^2$ are used. The $pK^+$ mass spectrum is shown in Fig. 1(c). No events are observed for $m_{pK^+}$ < 1.85 GeV/$c^2$. The bin size is $4\sigma_{pK^+}$, where $\sigma_{pK^+}$,

the detector resolution, varies from 1.6 to 4.2 MeV for $1.43 < m_{pK^+} < 1.85\,\text{GeV}/c^2$. The average $B^+ \to \Theta^{*++}(pK^+)\bar{p}$ signal efficiency in this region is $(17.0 \pm 0.2)\%$. A toy Monte Carlo[8] is used to calculate an upper limit in the presence of uncertainties on the efficiency and the number of expected background events. All the systematic errors except the background and the $B$-counting, contribute to the uncertainty on the efficiency (7.3%). The distribution of the combinatoric background events is obtained from a first-order polynomial fit to the $pK^+$ mass spectrum of data in the $m_{ES}$ sideband ($5.2 < m_{ES} < 5.26\,\text{GeV}/c^2$ and $|\Delta E| < 0.029\,\text{GeV}$) as well as $B^+ \to \eta_c(J/\psi)K^+ \to p\bar{p}K^+$ Monte Carlo events. The uncertainty on the background comes from the statistical error on the fit and the systematic error on the background. The resulting value of the upper limit at 90% confidence level (increased by the 1.1% error on $B$-counting) is $1.49 \times 10^{-7}$ for $1.43 < m(\Theta^{*++}) < 1.85\,\text{GeV}/c^2$.

## 4. Systematic Studies

The signal efficiencies are computed with $B^+ \to p\bar{p}K^+$ simulated events, reconstructed and selected using the same procedure as for the data. The statistical uncertainty on the efficiency leads to a 2.7(*1.1*[9])% systematic error. Corrections determined from data are applied to the efficiency calculation to account for the uncertainties in the tracking and PID performance. Systematic uncertainties assigned to each correction are 2.4% and 6% respectively. After all the corrections a $B^+ \to J/\psi(e^+e^-)K^+$ control sample is used to determine the residual differences in the efficiencies due to the cuts: 2% for the event shape and *2.5%* for the $m_{ES}$ and $\Delta E$ cuts. The $B$-backgrounds are varied by the uncertainties in their b.f.[7] resulting in a 1.2(*1.1*)% systematic error. The $B^+ \to p\bar{p}K^+$ b.f. measurement has a 4% systematic error due to the assumption of the signal shape under the $\eta_c$ and $J/\psi$. An uncertainty of 1.1% comes from the determination of the number of $B\bar{B}$ pairs.

## 5. Summary

In the data collected by the *BABAR* detector, the b.f. of $B^+ \to p\bar{p}K^+$ is measured to be $(6.7 \pm 0.9 \pm 0.6) \times 10^{-6}$ (total) and $(5.4 \pm 0.7 \pm 0.4) \times 10^{-6}$ (for $m_{p\bar{p}} < 2.85\,\text{GeV}/c^2$); at 90% confidence level the upper limit for $\mathcal{B}(B^+ \to \Theta^{*++}(pK^+)\bar{p})$ is set to be $1.49 \times 10^{-7}$ for $1.43 < m(\Theta^{*++}) < 1.85\,\text{GeV}/c^2$.